\documentclass[aps,pre,final]{revtex4}
\usepackage{times}
\usepackage{url}
\usepackage{amssymb}
\usepackage{amsmath}
\usepackage{bm}
\usepackage{graphicx}
\usepackage{xcolor}
\usepackage{setspace}
\usepackage{graphicx}
\usepackage{subfigure} 
\usepackage{algorithm}
\usepackage{algorithmic}
\usepackage{localmacros}
\usepackage{wrapfig}

\usepackage{tikz}
\usetikzlibrary{shapes,arrows,decorations.pathreplacing}
\tikzset{%
	factor/.style = {draw, thick, rectangle, minimum height = 1em, minimum width = 1em, fill = black},
	var/.style = {draw, thick, circle, minimum height = 2em, minimum width = 2em},
	link/.style = {coordinate}
}

\begin{document}

\title{Training Restricted Boltzmann Machines via the Thouless-Anderson-Palmer Free Energy}

\author{Marylou Gabri\'e$^{1,2}$, Eric W. Tramel$^1$ and
  Florent~Krzakala$^{1,3}$}

\affiliation{$^1$ Laboratoire de Physique Statistique, UMR 8550 CNRS,  Department of Physics, {\'E}cole Normale Sup{\'e}rieure and PSL  Research University, Rue Lhomond, 75005  Paris, France \\
  $^2$ International Centre for Fundamental Physics and its interfaces  at  Ecole normale supérieure, 75005  Paris, France \\
  $^3$ Sorbonne Universités, UPMC Univ Paris 06, UMR 8550, LPS,  F-75005, Paris, France}

\begin{abstract} 
  Restricted Boltzmann machines are undirected neural networks which
  have been shown to be effective in many applications, including
  serving as initializations for training deep multi-layer neural
  networks. One of the main reasons for their success is the existence
  of efficient and practical stochastic algorithms, such as
  contrastive divergence, for unsupervised training. We propose an
  alternative deterministic iterative procedure based on an improved
  mean field method from statistical physics known as the
  Thouless-Anderson-Palmer approach. We demonstrate that our algorithm
  provides performance equal to, and sometimes superior to, persistent
  contrastive divergence, while also providing a clear and easy to
  evaluate objective function. We believe that this strategy can be
  easily generalized to other models as well as to more accurate
  higher-order approximations, paving the way for systematic
  improvements in training Boltzmann machines with hidden units.
\end{abstract}

\date{\today}

\maketitle

\section{Introduction}
	\label{sec:intro}
	A restricted Boltzmann machine (RBM) \cite{smolensky1986chapter,Hinton2002} is a type of 
undirected neural network
with surprisingly many 
applications. This model has been used in problems as 
diverse as dimensionality reduction \cite{hinton2006reducing}, 
classification \cite{larochelle2008classification}, 
collaborative filtering \cite{salakhutdinov2007restricted}, 
feature learning \cite{coates2011analysis},
and topic modeling \cite{hinton2009replicated}. 
Also, quite remarkably, it has been shown that generative 
RBMs can be stacked into multi-layer 
neural networks, forming an initialization for discriminative
deep belief nets \cite{salakhutdinov2009deep}. Such deep architectures 
are believed to be crucial for 
learning high-order representations and concepts.

While the training procedure for RBMs can be written as a log-likelihood 
maximization, an exact implementation of this approach is computationally intractable
for all but the smallest models.
However, fast stochastic Monte Carlo methods, specifically contrastive divergence (CD)
\cite{Hinton2002} and persistent CD (PCD) \cite{Tieleman2008}, have made large-scale
RBM training both practical and efficient. These methods have popularized RBMs even
though it is not entirely clear why such approximate methods should work as well as 
they do.

In this paper, 
we propose an alternative deterministic
strategy for training RBMs, 
and neural networks with hidden units in general, 
based on the so-called \emph{mean field}, and 
\emph{extended mean field},
methods of statistical mechanics. 
This strategy has been used to train neural networks 
in a number of earlier works
\cite{Peterson1987,Hinton1989,Galland1993,Kappen1998,Welling2002}. In fact, for entirely 
visible networks, the use of adaptive cluster expansion mean field methods have lead to 
spectacular results in learning Boltzmann machine representations 
\cite{cocco2009neuronal,cocco2011adaptive,weigt2009identification}.

However, unlike these fully visible models, the hidden units of the RBM must be 
taken into account during the training procedure.
In 2002, Welling and Hinton \cite{Welling2002} presented 
a similar deterministic mean field learning algorithm for general Boltzmann machines with hidden units, considering 
it \emph{a priori} as a potentially efficient extension of CD. 
In 2008, Tieleman \cite{Tieleman2008} tested the method in detail for RBMs and found it 
provided poor performance when compared to both CD and PCD. 
In the wake of these two papers, little inquiry has been made in this
direction, with the apparent consensus being that the 
deterministic mean field approach is ineffective for RBM training.

Our goal is to challenge this consensus by going beyond na{\"i}ve mean
field,
a mere first-order approximation, by introducing
second-, and possibly third-, order terms. 
In principle, it is even possible to extend the approach to 
arbitrary order.
Using this extended mean field approximation, commonly known as 
the Thouless-Anderson-Palmer \cite{Thouless1977} approach in statistical physics, 
we find that RBM training performance is significantly improved over the
na{\"i}ve mean field approximation
and is even comparable to PCD. 
The clear and easy to evaluate objective function, along with the
extensible nature of the approximation, paves the way for systematic 
improvements in learning efficiency.

\section{Training restricted Boltzmann machines}
	\label{sec:basic}
	A restricted Boltzmann machine,
which can be viewed as a two layer
undirected bipartite neural network,
is a specific case of an energy based model
wherein a 
layer of visible units are 
fully
connected to a layer of hidden units. 
Let us denote the binary visible and hidden units, 
indexed by $i$ and $j$ respectively, as $v_i$ and $h_j$.
The energy of a given state, 
$\mathbf{v}=\{v_i\}$, $\mathbf{h}=\{h_j\}$, of the RBM is given by
\begin{equation}
 E(\mathbf{v},\mathbf{h})=- \sum_{i} a_i v_i - \sum_{j} b_j h_j -\sum_{i,j}v_i w_{ij} h_j  ,
\end{equation}
where $w_{ij}$ are the entries of the matrix specifying the weights, 
or \emph{couplings}, 
between the visible and hidden units, and $a_i$ and $b_j$ are the biases, 
or the \emph{external fields} in the language of statistical physics, 
of the visible and hidden units, respectively. 
Thus, the set of parameters $\{w_{ij},a_i,b_j\}$ define the RBM model.

The joint probability distribution over the visible and hidden units is given 
by the Gibbs-Boltzmann measure 
$P(\mathbf{v},\mathbf{h})={Z}^{-1}e^{-E(\mathbf{v},\mathbf{h})}$, 
where $Z=\sum_{\mathbf{v},\mathbf{h}}e^{-E(\mathbf{v},\mathbf{h})}$ 
is the normalization constant known as the \emph{partition function} in physics. 
For a given data point,
represented by $\mathbf{v}$, the marginal of the RBM is
calculated as
$P(\mathbf{v})=\sum_{\mathbf{h}}P(\mathbf{v},\mathbf{h})$. 
Writing this marginal of $\mathbf{v}$ in terms of its log-likelihood 
results in the difference
\begin{equation}
\mathcal{L}=\ln P(\mathbf{v})= -F^{c}(\mathbf{v}) + F,
\end{equation}
where $F=-\ln Z $ is the \emph{free energy} of the RBM, 
and $F^c(\mathbf{v})=-\ln(\sum_{\mathbf{h}}e^{-E(\mathbf{v},\mathbf{h})})$ can
be interpreted as a free energy as well, but with visible units fixed to the training data point $\mathbf{v}$. Hence, $F^c$ is referred to as the \emph{clamped} free energy. 

One of the most important features of the RBM model is that $F^c$ can be easily computed
as $\mathbf{h}$ may be summed out analytically since the hidden units are conditionally 
independent of the visible units, owing to the RBM's bipartite structure.
However, calculating $F$ 
is computationally intractable since the number of possible states to
sum over scales 
combinatorially
with the 
number of units in the model. 
This complexity frustrates the exact computation of the 
gradients of the log-likelihood 
needed in order to train the RBM parameters via gradient ascent. 
Monte Carlo methods for RBM training rely on the 
observation that $\frac{\partial F}{\partial w_{ij}}=P(v_i=1,h_j=1)$,
which can be simulated at a lower computational cost. 
Nevertheless, drawing independent samples from the model 
in order to approximate this derivative is itself computationally 
expensive and often approximate sampling 
algorithms, such as CD or PCD, are used instead.

%
%
%

\section{Extended mean field theory of RBMs}
	\label{sec:theory}
	\label{GY}
Here, we present a physics-inspired tractable estimation of
the free energy $F$ of the RBM. This approximation is based on a high
temperature expansion of the free energy derived by 
Georges and Yedidia in the context of spin glasses \cite{Georges1999} 
following the pioneering works of \cite{Thouless1977,Plefka1982}. We refer
the reader to \cite{opper2001advanced} for a review of this topic. 

To apply the Georges-Yedidia expansion to the RBM free energy,
we start with a general energy based model which possesses 
arbitrary couplings $w_{ij}$ between undifferentiated binary spins
$s_{i}\in \{0,1\}$, 
such that the energy of the Gibbs-Boltzmann measure on the
configuration $\mathbf{s} = \{s_i\}$ is defined by
$E(\mathbf{s})=-\sum_{i}a_i s_i
-\sum_{i,j} w_{ij} s_i s_j$. 
We also restore the role of the
temperature, usually set to 1 in most energy based models, by multiplying
the energy functional in the Boltzmann weight by the inverse temperature
$\beta$. 


Next, we apply a Legendre transform to the free energy, a standard procedure in
statistical physics, by first writing the free energy as 
a function of a newly introduced auxiliary external field $\mathbf{q}=\{q_i\}$,
$ -\beta F[\mathbf{q}] = \ln \sum_{\mathbf{s}} \mathrm{e} ^{- \beta E(\mathbf{s})+ \beta \sum_i q_i s_i}$.
This external field will be eventually set to the value 
$\mathbf{q}=\mathbf{0}$ 
in order to recover the true free energy. The Legendre transform $\Gamma$ 
is then given as a function of the conjugate variable $\mathbf{m}=\{m_i\}$
by maximizing over $\mathbf{q}$,
\begin{equation}
-\beta \Gamma[\mathbf{m}]
	= -\beta \max_{\mathbf{q}}[F[\mathbf{q}]+\sum_i q_i m_i] 
	= -\beta (F[\mathbf{q}^*[\mathbf{m}]]+\sum_i q_i^*[\mathbf{m}] m_i),
\end{equation}
where the maximizing auxiliary field 
$\mathbf{q}^*[\mathbf{m}]$,
a function of the conjugate variables,
is the inverse function of 
$\mathbf{m}[\mathbf{q}] \equiv -\frac{dF}{d\mathbf{q}}$. 
Since the derivative $\frac{dF}{d\mathbf{q}}$ is exactly 
equal to $ - \langle \mathbf{s} \rangle$, where the 
operator $\langle\cdot\rangle$ refers to the average configuration under
the Boltzmann measure, the conjugate variable 
$\mathbf{m}$ is in fact the equilibrium \emph{magnetization} vector 
$\langle \mathbf{s} \rangle$. Finally, we observe that 
the free energy is also the \emph{inverse}
Lengendre transform of its Legendre transform at $\mathbf{q}=\mathbf{0}$,
	\begin{equation}
	\label{inverseLg}
	-\beta F=-\beta F[\mathbf{q}=\mathbf{0}]=\beta \min_{\mathbf{m}}[\Gamma[\mathbf{m}]] =-\beta \Gamma[\mathbf{m^*}],
	\end{equation} 
where $\mathbf{m}^*$ minimizes $\Gamma$.

Following \cite{Plefka1982,Georges1999},
this formulation allows us to perform a high temperature expansion of 
$A(\beta,\mathbf{m})\equiv -\beta \Gamma[\mathbf{m}]$ 
around $\beta=0$  at fixed $\mathbf{m}$,
\begin{equation}
A(\beta,\mathbf{m})
	=A(0,\mathbf{m}) + \beta \left.\frac{\partial A(\beta,\mathbf{m})}{\partial \beta}\right|_{\beta=0}+\frac{\beta^2}{2}\left.\frac{\partial^2 A(\beta,\mathbf{m})}{\partial \beta^2}\right|_{\beta=0} + \cdots,
\end{equation}
where the dependence on $\beta$ of the product $\beta \mathbf{q}$ must carefully 
be taken into account. At infinite temperature, $\beta = 0$,
the spins decorrelate, causing the average value of an
arbitrary product of spins to equal the product of 
their local magnetizations; a useful property. 
Accounting for binary spins taking values in $\{0,1\}$, one obtains 
the following expansion
\begin{align}
\label{Gamma} 
-\beta \Gamma (\mathbf{m})
=&-\sum_i \left[ m_i \ln m_i
	+(1-m_i)\ln(1-m_i) \right] + \beta \sum_i a_i m_i  +\beta \sum_{(i,j)} w_{ij} m_i m_j \nonumber\\
&+\frac{\beta^2}{2} \sum_{(i,j)} w_{ij}^2
	(m_i-m_i^2) (m_j-m_j^2) \nonumber \\
&+ \frac{2 \beta^3}{3}\sum_{(i,j)} w_{ij}^3
	(m_i-m_i^2) \left(\frac{1}{2}-m_i\right)
	(m_j-m_j^2)\left(\frac{1}{2}-m_j\right) \nonumber \\
&+ \beta^3 \sum_{(i,j,k)}
	w_{ij}w_{jk}w_{ki}(m_i-m_i^2) (m_j-m_j^2)(m_k-m_k^2) +
	\cdots 
\end{align}
The zeroth-order term corresponds to the entropy of non-interacting spins with 
constrained magnetizations values.  Taking this expansion
up to the first-order term, we recover 
the standard na{\"i}ve mean field theory. 
The second-order term is known as the 
\emph{Onsager reaction term} in the TAP equations \cite{Thouless1977}. 
The higher orders terms are systematic corrections which were first derived 
in \cite{Georges1999}.

Returning to the RBM notation and truncating the expansion
at second-order for the remainder of the theoretical discussion, we have
\begin{align}
\label{GammaRBM}
\Gamma(\mathbf{m^v},\mathbf{m^h})
&\simeq S(\mathbf{m}^v,\mathbf{m}^h) - \sum_i a_i m^v_i - \sum_j b_j m^h_j 
\nonumber\\
&\quad - \sum_{i,j} w_{ij} m^v_i m^h_j   + \frac{w_{ij}^2}{2} (m^v_i-{m^v_i}^2)(m^h_j-{m^h_j}^2), 
\end{align}
where $S$ is the entropy contribution,  
$\mathbf{m^v}$ and $\mathbf{m^h}$ are
introduced to denote the magnetization of the visible 
and hidden units, and $\beta$ is set equal to 1. 
Eq. \eqref{GammaRBM} can be viewed as a 
\emph{weak coupling} expansion in $w_{ij}$. 
To recover an estimate of the RBM free energy, Eq. \eqref{GammaRBM}
must be minimized with respect to its arguments, as in Eq. \eqref{inverseLg}. 
Lastly, by writing the stationary condition $\frac{d\Gamma}{d\mathbf{m}}=\mathbf{0}$,
we obtain the self-consistency constraints on the magnetizations. 
For instance, at second-order we obtain the following constraint on
the visible magnetizations,
\begin{equation}
\label{self}
m^{v}_i  
\simeq \mathrm{sigm}\left[ a_i + \sum_{j} w_{ij}m^{h}_j - w_{ij}^2 \left(m_i^{v}-\frac{1}{2}\right) \left(m^h_j-{m^h_j}^2\right)  \right],
\end{equation}
where $\sigm[x] = (1+e^{-x})^{-1}$ is a logistic sigmoid function.
A similar constraint must be satisfied for the hidden units, as well. 
Clearly, the stationarity condition for $\Gamma$ obtained at order $n$ 
utilizes terms up to the $n^{th}$ order within the sigmoid argument of these 
consistency relations. Whatever the order of the approximation, the magnetizations 
are the solutions of a set of non-linear coupled equations of the 
same cardinality as the number of units in the model. 
Finally, provided we can define a procedure to efficiently derive the value of the magnetizations 
satisfying these constraints, 
we obtain an extended mean field approximation of 
the free energy which we denote as $\Femf$. 



\section{RBM evaluation and unsupervised training with EMF}
\subsection{An iteration for calculating $\Femf$} 
Recalling the log-likelihood of the RBM, 
$\mathcal{L}=-F^{c}(\mathbf{v}) + F$, we have shown that a tractable
approximation of $F$, $\Femf$, is obtained via a weak coupling
expansion so long as one can solve the coupled system of equations over the magnetizations shown in Eq. \eqref{self}. 
In the spirit of iterative belief propagation \cite{opper2001advanced}, 
we propose that these self-consistency relations can serve as update rules for
the magnetizations within an iterative algorithm.
In fact, the convergence of this
procedure has been rigorously demonstrated in the context of random
spin glasses \cite{bolthausen2014iterative}. 
We expect that these convergence properties will remain present
even for real data.
The iteration over the self-consistency relations for both the 
hidden and visible magnetizations can be written using the time
index $t$ as
\begin{align}
m^{h}_j[t+1]
&\gets \mathrm{sigm}\left[ b_j + \sum_{i} w_{ij}m^{v}_i[t] - w_{ij}^2 
			\left(m_j^h[t]-\frac{1}{2}\right) 
			\left(m^v_i[t]-(m^v_i[t])^2\right)  \right], 
			\tag{9,~10}\label{update}\\
m^{v}_i[t+1]
&\gets \mathrm{sigm}\left[ a_i + \sum_{j} w_{ij}m^{h}_j[t+1] - w_{ij}^2 
			\left(m_i^{v}[t]-\frac{1}{2}\right) 
			\left(m^h_j[t+1]-(m^h_j[t+1])^2\right)  \right],\nonumber
\end{align}
\addtocounter{equation}{2}%
where the time indexing follows from careful application of \cite{bolthausen2014iterative}.
The values of $\mathbf{m}^v$ and $\mathbf{m}^h$ minimizing 
$\Gamma(\mathbf{m}^v,\mathbf{m}^h)$, and thus providing the value of
$\Femf$, are obtained by running Eqs. \eqref{update} until
they converge to a fixed point. We note that while we present an iteration to
find $\Femf$ up to second-order above, third-order terms can easily be introduced 
into the procedure.
%

\subsection{Deterministic EMF training}
\label{EMFtrainingprinciples}
By using the EMF estimation of $F$, and the iterative algorithm detailed in
the previous section to calculate it, 
it is now possible to estimate to gradients of the log-likelihood
used for unsupervised training of the RBM model by substituting $F$ with
$\Femf$. We note that the deterministic iteration we propose for estimating $F$ 
is in stark contrast with the stochastic sampling procedures utilized in CD and
PCD to the same end.
For instance, the gradient ascent update of weight $w_{ij}$ is approximated as
\begin{equation}
\Delta w_{ij} 
	\propto \frac{\partial \mathcal{L}}{\partial w_{ij}}
	\simeq-\frac{\partial F^c}{\partial w_{ij}}+\frac{\partial \Femf}{\partial w_{ij}},
\end{equation}
where $\frac{\partial \Femf}{\partial w_{ij}}$ can be computed by differentiating
Eq. \eqref{GammaRBM} at fixed
$\mathbf{m}^v$ and $\mathbf{m}^h$ and computing the value of these derivatives at 
the fixed points of Eqs. \eqref{update} obtained from the iterative procedure. 
The gradients with respect to the visible and hidden biases can be derived similarly. 
Interestingly, 
$\frac{\partial \Femf}{\partial a_i}$ and
$\frac{\partial \Femf}{\partial b_j}$ are merely the fixed-point magnetizations
of the visible and hidden units, $m^v_i$ and $m^h_j$, respectively.

\emph{A priori}, 
the training procedure sketched above can be used at any order of 
the weak coupling expansion. 
The training algorithm introduced in \cite{Welling2002}, which was shown
to perform poorly for RBM training in \cite{Tieleman2008}, can be recovered by retaining
only the first-order of the expansion when calculating $\Femf$.
By taking $\Femf$ to second-order, we expect that training 
efficiency and performance will be greatly improved over \cite{Welling2002}.
In fact, including the third-order term in the training algorithm
is just as easy as including the second-order one, 
due to the fact that the particular structure of the
RBM model does not admit triangles in its corresponding factor graphs.
Although the third-order term in Eq. \eqref{Gamma} does 
include a sum over distinct pairs of units, as well as a sum over coupled 
triplets of units, such triplets
are excluded by the bipartite structure of the RBM. 
However, coupled \emph{quadruplets} do contribute to the fourth-order term
and therefore fourth- and higher-order approximations 
require much more expensive computations \cite{Georges1999}, though
it is possible to utilize adaptive procedures as
in \cite{cocco2011adaptive}.




\section{Numerical experiments}
	\label{sec:experiments}

\subsection{Experimental framework}
To evaluate the performance of the proposed deterministic EMF 
RBM training algorithm, we perform a number of numerical experiments
over two separate datasets and compare these results with both 
CD-1 and PCD. 
We first use the MNIST dataset of labeled handwritten
digit images \cite{LeCun1998}. The dataset is split between $60\,000$
training images and $10\,000$ test images. Both subsets
contain approximately the same fraction of the ten digit classes (0 to
9). Each image is comprised of $28\times28$ pixels taking values 
in the range $[0,255]$. 
The MNIST dataset was binarized by setting all non-zero pixels to 1 
in all experiments, with the exception of one experiment 
in which we train on a version of the MNIST dataset rescaled to $[0,1]$.

Second, we use
the $28\times28$ pixel version of the Caltech 101 Silhouette dataset
\cite{Marlin2010}. Constructed from
the Caltech 101 image dataset, the silhouette dataset consists of 
black regions of the
primary foreground scene object on a white background. The images are labeled
according to the object in the original picture, of which there
are 101 unevenly represented object labels. The dataset is split between a training
($4\,100$ images), a validation ($2\,264$ images),
and a test ($2\,304$ images) sets.

For both datasets, the RBM models require 784 visible units. 
Following previous studies evaluating RBMs on these datasets,
we fix the number of RBM hidden units to 500 in all our experiments. 
During training, we adopt the mini-batch learning procedure for
gradient averaging, 
with 100 training points per batch for MNIST and 256 training points 
per batch for Caltech 101 Silhouette.

We test the EMF learning algorithm presented in Section
\ref{EMFtrainingprinciples} in various settings. First,
we compare implementations utilizing the first-order (MF),
second-order (TAP2), and third-order (TAP3) approximations of $F$. 
Higher orders were not considered due to their greater complexity.
Next, we investigate training quality when the self-consistency
relations on the magnetizations were not converged when calculating
the derivatives of $\Femf$,
but instead iterated for only a small fixed (3) number of iterations,
an approach similar to CD-1.
Furthermore, we also evaluate a
``persistent'' version of our algorithm, similar to \cite{Tieleman2008}. 
In this implementation, the magnetizations of a set of points, dubbed fantasy
particles, are updated and maintained throughout the training in order to estimate
$F$. 
This persistent procedure takes advantage of the fact that the RBM-defined Boltzmann measure changes only slightly between
training epochs.
Convergence to the new fixed point magnetizations at each epoch 
should therefore be sped up by initializing with the converged state from the previous update. Our final experiments
consist of persistent training algorithms using 3 iterations of the 
magnetization self-consistency relations (P-MF, P-TAP2 and P-TAP3) and 
one persistent training algorithm using 30 iterations (P-TAP2-30) for comparison. 

Lastly, we evaluate RBM training for the rescaled, non-binarized, 
MNIST dataset (P-TAP2 raw). This experiment is designed to mimic the
pre-training of a second stacked RBM of a deep belief net.
In this setting, the training data for the second RBM consists of non-binary magnetizations 
derived from the hidden units of first RBM operating on the true binary training
data \cite{hinton2006reducing}. 
For this experiment, 
EMF iterations were used to estimate
both the clamped term as well as the free energy term in the computation
of the log-likelihood gradients.



For comparison,
we also train RBM models using CD-1, 
following the prescriptions of \cite{Hinton2010},
and PCD, as implemented by Tieleman \cite{Tieleman2008}. Given that our goal is
to compare RBM training approaches rather than achieving the best
\emph{possible} training across all free parameters, 
neither momentum nor adaptive learning rates were
included in any of the implementations tested. However,
we do employ a weight decay
regularization in all our trainings to keep weights small; 
a necessity for the weak coupling expansion on which the EMF relies.
When comparing learning procedures on the same plot, all 
free parameters of the training 
(e.g. learning rate, weight decay, etc.) 
were set identically. 
All results are presented as averages
over 10 independent trainings with standard deviations reported as error bars.

\subsection{Relevance of the EMF log-likelihood}

\begin{figure}[t!]
	  \begin{center}
	    \includegraphics[width=0.9\textwidth]{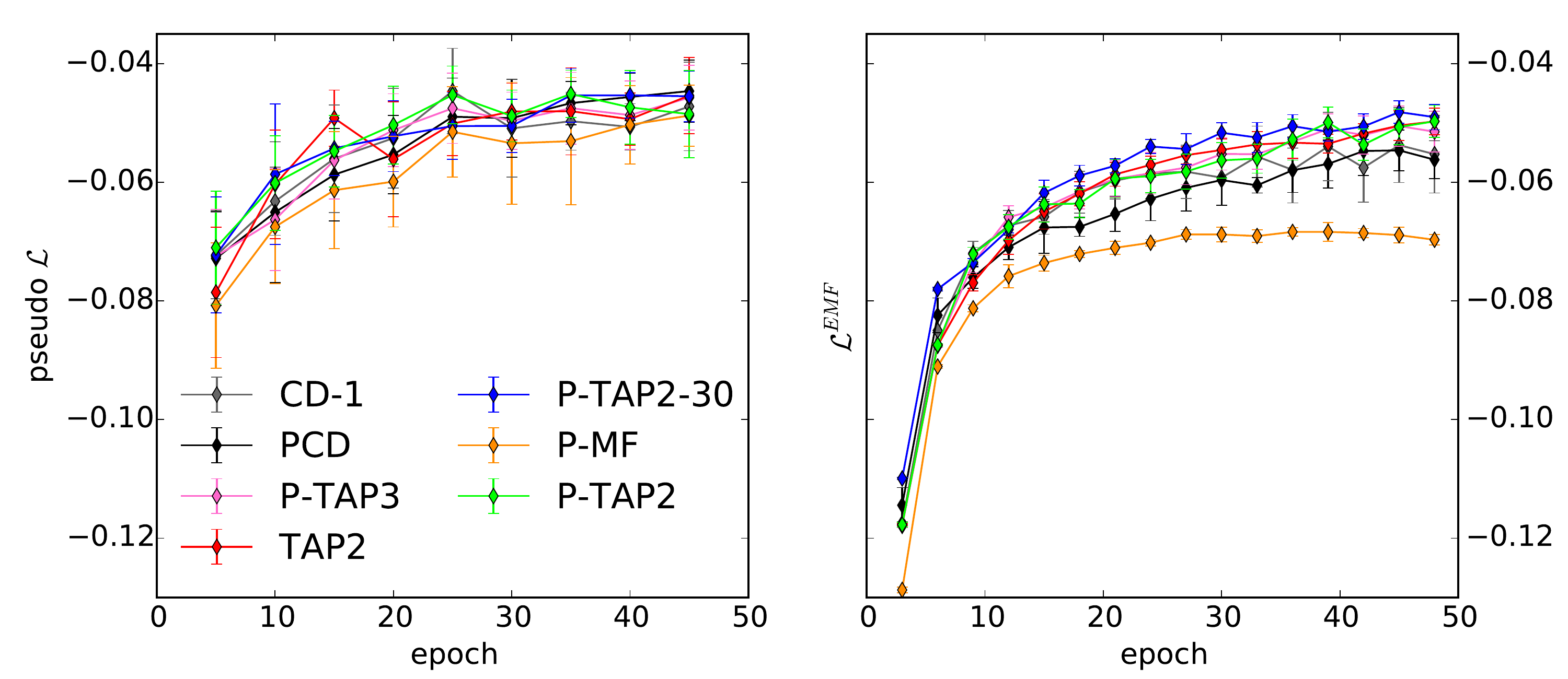}
	   \caption{Estimates of the per-sample log-likelihood over the
             MNIST test set, normalized by the total number
             of units, as a function of the number of training epochs
             for a 500 hidden unit RBM. 
             The results for the different
             training algorithms are plotted in different colors with the
             same color code used for both panels. 
             {\bf Left panel :} 
              Pseudo log-likelihood estimate. The difference between EMF algorithms and contrastive divergence algorithms is minimal. 
             {\bf Right panel :} 
              EMF log-likelihood estimate at $2^{nd}$ order. 
              The improvement from MF to TAP is clear. Perhaps 
              reasonably, TAP demonstrates an advantage over CD and PCD. 
              Notice how the second-order EMF approximation of $\mathcal{L}$
              provides less noisy estimates, at a lower computational cost.}
	   \label{MNISTlikelihood}
	  \end{center}
\end{figure}	


\begin{figure}
\center
\begin{tikzpicture}[auto,thick, node distance = 3em, >=triangle 45,scale=1]
    \node[anchor=north west,inner sep=0] at (0,4.4) 
        {\includegraphics[width=5in]{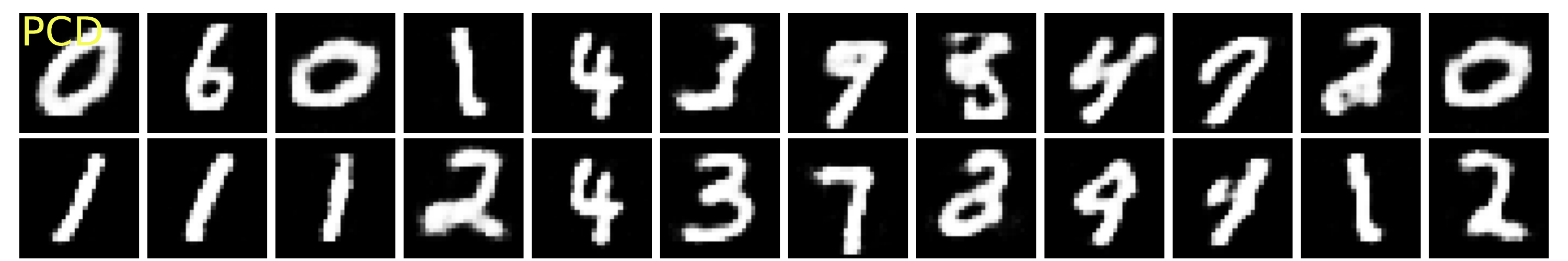}};
    \node[anchor=north west,inner sep=0] at (0,2.2) 
        {\includegraphics[width=5in]{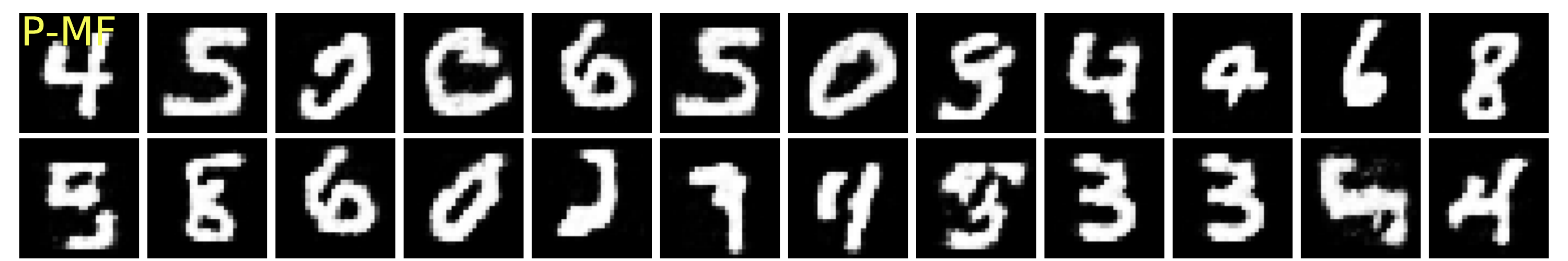}};
    \node[anchor=north west,inner sep=0] at (0,0) 
    	{\includegraphics[width=5in]{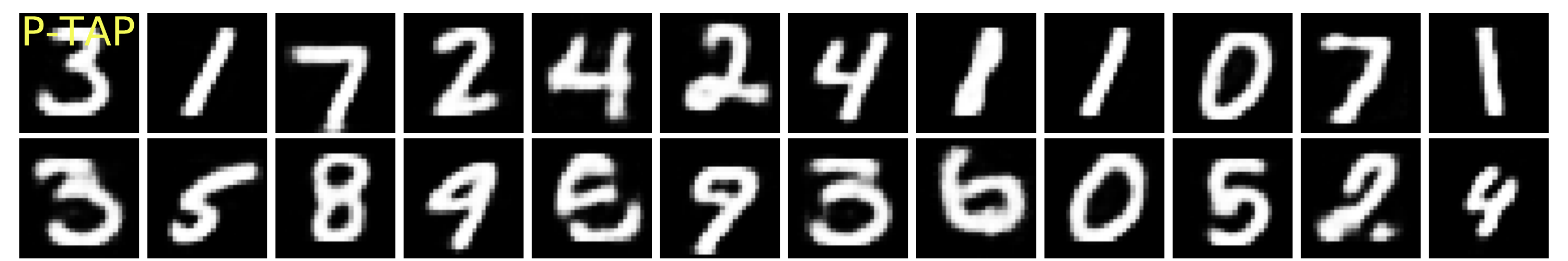}};
\end{tikzpicture}
\caption{
  Fantasy particles generated by a 500 hidden unit RBM after
  50 epochs of training on the MNIST dataset with PCD 
  ({\bf top two rows}), P-MF ({\bf middle two rows}) and P-TAP2 ({\bf bottom two rows}). 
  These fantasy particles represent typical samples generated by the trained RBM 
  when used as a generative prior for handwritten numbers. 
  The samples generated by P-TAP2 are of similar subjective quality, 
  and perhaps slightly preferable, to those generated by PCD, 
  while certainly preferable to those generated by P-MF.}
	   \label{FantasyParticles}
\end{figure}

\begin{figure}[t!]
	  \begin{center}
	    \includegraphics[width=0.9\textwidth]{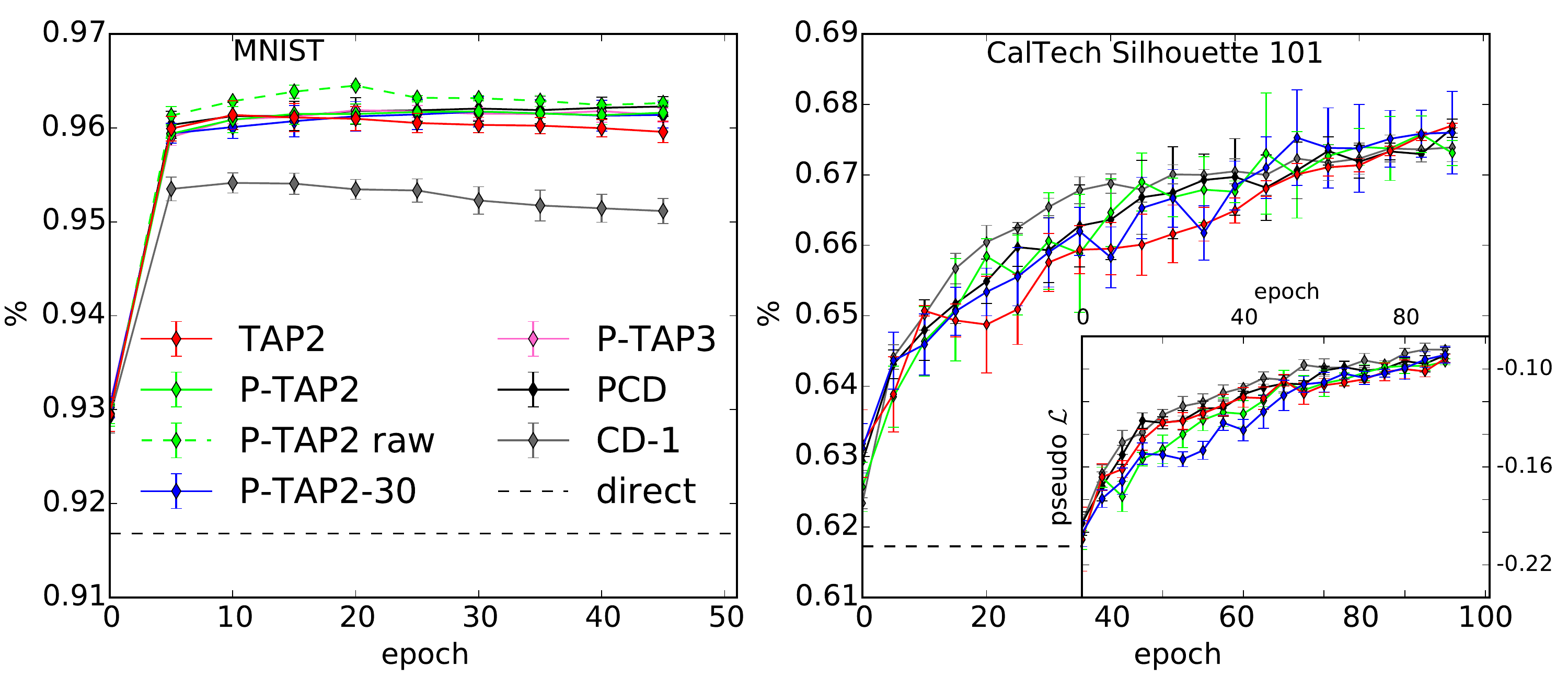}
	   \caption{%
       Test set classification accuracy for the MNIST ({\bf left}) 
       and Caltech 101 Silhouette ({\bf right}) datasets using 
       logistic regression on the hidden unit marginal probabilities 
       as a function of the number of epochs used to train a 500 hidden unit RBM.  
       As a baseline comparison, the classification accuracy of logistic 
       regression performed directly on the data is given as a black dashed line.
       Here, P-TAP2 raw refers to the P-TAP2 training algorithm performed on
       the rescaled, non-binarized, version of the MNIST dataset.
       The results for the different 
       training algorithms are displayed in different colors, with the same color 
       code being used in both panels. ({\bf Right inset:}) Pseudo log-likelihood
       over training epochs for the Caltech 101 Silhouette dataset.
       \label{Classification}}	   
	  \end{center}
\end{figure}

Our first observation is that the implementations of the EMF training algorithms
are not overly belabored.
The free parameters relevant for the PCD and CD-1 procedures were found to be equally 
well suited for the EMF training algorithms. 
In fact, as shown in the left panel of Fig. \ref{MNISTlikelihood}, and the
right inset of Fig. \ref{Classification},
the ascent of the pseudo log-likelihood over training epochs is very similar between the EMF
training methods and both the CD-1 and PCD trainings.

Interestingly, for the Caltech 101 Silhouettes dataset, it seems that 
the persistent algorithms tested have difficulties in 
ascending the pseudo-likelihood in the first 
epochs of training. This contradicts the common belief that persistence yields 
more accurate
approximations of the likelihood gradients. 
The complexity of the training set, 101 classes 
unevenly represented over only $4\,100$ training points, might explain this unexpected behavior.
The persistent fantasy particles all converge to similar
non-informative blurs in the earliest training epochs with many epochs
being required to resolve the particles to a distribution of values
which are informative about the pseudo log-likelihood.

Examining the fantasy particles also gives an idea of the performance
of the RBM as a generative model. In Fig. \ref{FantasyParticles}, 24
randomly chosen fantasy particles from the $50^{th}$ epoch of training
with PCD, P-MF, and P-TAP2 are displayed. 
The RBM trained with PCD generates recognizable digits, yet
the model seems to have trouble generating several
digit classes, such as 3, 8, and 9.
The fantasy particles extracted from a P-MF training are of
poorer quality, with half of the drawn particles featuring non-identifiable
digits. The P-TAP2 algorithm, however, appears to provide
qualitative improvements.
All digits can be visually discerned, with visible
defects found only in two of the particles.
These particles seem to indicate that it is indeed possible to efficiently
persistently train an RBM without converging on the fixed point of the magnetizations.

The relevance of the EMF log-likelihood for RBM training
is further confirmed in the right panel of Fig. \ref{MNISTlikelihood},
where we observe that both CD-1 and PCD ascend the second-order
EMF log-likelihood, \emph{even though they are not explicitly constructed
to optimize over this objective}.
As expected, the persistent TAP2 algorithm with 30 iterations of the magnetizations
(P-TAP2-30) achieves the best maximization of
$\mathcal{L}^{EMF}$. 
However, P-TAP2, with only 3 iterations of the magnetizations, achieves very similar
performance, perhaps making it preferable when a faster training algorithm is desired.
Moreover, we note that although P-TAP2 demonstrates
improvements with respect to the P-MF, the P-TAP3 does not yield
significantly better results than P-TAP2. This is perhaps not
surprising since the third order term of the EMF expansion consists of
a sum over as many terms as the second order, but at a smaller order in
$\{w_{ij}\}$. 

\subsection{Classification task performance}
We also evaluate these RBM training algorithms from the perspective of
supervised classification.
An RBM can be interpreted as a deterministic function mapping the
binary visible unit values to the real-valued hidden unit magnetizations. 
In this case, the hidden unit magnetizations
represent the contributions of some learned features. Although
no supervised fine-tuning of the weights is implemented, we tested
the quality of the features learned by the different training
algorithms by their usefulness in classification tasks. 
For both datasets, a logistic regression classifier was
calibrated with the hidden units magnetizations mapped from the
labeled training images using the \texttt{scikit-learn} toolbox
\cite{scikit-learn}. We purposely avoid using more
sophisticated classification algorithms in order to place emphasis on
the quality of the RBM training, not the classification method.

In Fig. \ref{Classification}, we see that the MNIST classification accuracy 
of the RBMs trained with the P-TAP2 algorithms is roughly equivalent
with that obtained when using PCD training, while CD-1 training yields
markedly poorer classification accuracy.
The slight decrease in
performance of CD-1 and TAP2 along as the training epochs increase
might be emblematic of over-fitting by the non-persistent 
algorithms, although no decrease in the EMF test set log-likelihood was observed. 
We note that
the classification accuracy for the RBM trained on the rescaled MNIST
data (P-TAP2 raw) is marginally better than the other tested approaches, 
which implies that the training the RBM on real-valued data was
successful. Consequently, even if our algorithm is designed for
binary visible units, it can be used equally with real visible variables
by treating them as magnetizations, just as is commonly done with
the CD-1 algorithm.

Finally, for the Caltech 101 Silhouettes dataset, the classification task,
shown in the right panel of Fig. \ref{Classification}, is 
much more difficult \emph{a priori}. 
Interestingly, the persistent algorithms do
not yield better results on this task. However, 
we observe that the 
performance of 
deterministic EMF RBM training is at least comparable with
both CD-1 and PCD.

\section{Conclusion}
	\label{sec:conclusion}
	We have presented a method for training RBMs based on an extended mean field approximation. 
Although a na{\"i}ve mean field learning algorithm had already been designed for RBMs,
and judged 
unsatisfactory \cite{Welling2002,Tieleman2008}, we have shown that extending beyond 
the na{\"i}ve mean field to include terms of second-order and above 
brings significant improvements over the first-order approach and 
allows for practical and efficient deterministic RBM training with 
performance comparable to the stochastic CD and PCD training algorithms.
A demo file, with an implementation of our algorithm, is provided
online\footnote{It can be downloaded in the SPHINX group software
webpage \url{http://www.lps.ens.fr/\~krzakala/WASP.html}.}.

The extended mean field theory also provides an estimate of the RBM 
log-likelihood which is easy 
to evaluate and thus enables practical monitoring of the progress of unsupervised learning 
throughout the training epochs. 
Furthermore, training on real-valued magnetizations is 
theoretically well-founded within the presented approach and was shown to be 
successful in experimentation. These results pave the way for many possible extensions. 
For instance, it would be quite straightforward to apply the same kind of expansion 
to Gauss-Bernoulli RBMs, as well as to multi-label RBMs.

The extended mean field approach might also be
used to learn stacked RBMs jointly, rather than separately, as is done in both
deep Boltzmann machine and deep belief network pre-training, a strategy 
that has shown some promise \cite{goodfellow2013joint}. In fact, the 
approach can be generalized even to non-restricted Boltzmann machines with hidden variables
with very little difficulty. 
Another interesting possibility would be to make use of 
higher-order terms in the series expansion using adaptive cluster methods such as 
those used in \cite{cocco2011adaptive}. We believe our results show that the extended mean
field approach, and in particular the Thouless-Anderson-Palmer one, may be a good starting 
point to theoretically analyze the performance of RBMs and deep belief networks.

 \section*{Acknowledgment}
 The research leading to these results has received funding from the
 European Research Council under the European Union's $7^{th}$
 Framework Programme (FP/2007-2013/ERC Grant Agreement
 307087-SPARCS). We thank Lenka Zdeborov\'a for numerous discussion.

\bibliographystyle{IEEEtran}
\small{
	\bibliography{BiblioGY}	
}

\end{document}